\documentclass[referee,useAMS,usenatbib,12pt]{mn2e}
\usepackage{graphicx,graphics,amsmath,amssymb,mathrsfs}
\newcommand{\nin}{\noindent}
\newcommand{\dmath}{\mbox{${\mathrm d}$}}

\newcommand{\der}[2]{\frac{\dmath{#1}}{\dmath{#2}}}

\newcommand{\dder}[2]{\frac{\dmath\,^2{#1}}{\dmath{#2}^2}}

\newcommand{\pom}{\dot{\varpi}}

\newcommand{\siga}[2]{\Sigma_{#1}^{#2}}
\newcommand{\bfpar}[3]{{\bf{#1}}_{#2}^{#3}}
\newcommand{\vtilde}{{\tilde{v}}_{\phi}}

\newcommand{\rmi}{{\rm i}}

\newcommand{\rme}{{\rm e}}

\newcommand{\intglwr}[1]{\int^{\infty}_{#1}}

\newcommand{\intgl}{\int_{-\infty}^{\infty}}

\newcommand{\sss}[1]{{\scriptscriptstyle{#1}}}
\newcommand{\rmpar}[1]{{\rm {#1}}}
\newcommand{\sigd}[1]{\Sigma_{\rmpar{d}}^{#1}}
\newcommand{\md}{M_{\rmpar{d}}}

\def\apj{{Astroph.\@ J. }}
\def\mnras{{Mon.\@ Not.\@ Roy.\@ Ast.\@ Soc.}}
\def\aap{{Astron.\@ Astrophys.}}
\def\aj{{Astron.\@ J.}}

\def\apjl{{Astrophysical.\@ J. {\rm Letters}}}

\begin{document}
\title{Modal analysis of gravitational instabilities in nearly Keplerian, counter-rotating collisionless discs}
\author[Gulati \& Saini]{Mamta Gulati$^{1,3}$, Tarun Deep Saini$^{2,4}$ \\
  $^{1}$ Indian Institute of Science Education and Research Mohali, Chandigarh, 560 012, India\\
  $^{2}$ Indian Institute of Science, Bangalore 560 012, India\\	
  $^{3}$ mgulati@iisermohali.ac.in\,,
  $^{4}$ tarun@physics.iisc.ernet.in\,.
}
\maketitle

\begin{abstract}
We present a modal analysis of instabilities of counter-rotating, self-gravitating 
collisionless stellar discs, using the recently introduced modified WKB formulation 
of spiral density waves for collisionless systems (Gulati \& Saini). The discs are 
assumed to be axisymmetric and in coplanar orbits around a massive object at the 
common center of the discs. The mass in both discs is assumed to be much smaller 
than the mass of the central object. For each disc, the disc particles are assumed 
to be in near circular orbits. The two discs are coupled to each other gravitationally. 
The perturbed dynamics of the discs evolves on the order of the precession time scale 
of the discs, which is much longer than the Keplerian time scale. We present results 
for the azimuthal wave number $m=1$ and $m=2$, for the full range of disc mass ratio 
between the prograde and retrograde discs. The eigenspectra are in general complex, 
therefore all  eigenmodes are unstable. Eigenfunctions are radially more compact for 
$m = 1$ as compared to $m = 2$. Pattern speed of eigenmodes is always prograde with 
respect to the more massive disc. The growth rate of unstable modes increases with 
increasing mass fraction in the retrograde disc, and decreases with $m$; therefore 
$m=1$ instability is likely to play the dominant role in the dynamics of such systems. 
\end{abstract}

\section{Introduction}\label{intro}
Observations of galactic nuclei are limited by the resolution of present day telescopes. 
Of the few galaxies for which such observations are available, double peak stellar 
distribution has been observed in galaxies M$31$, a spiral galaxy, and NGC$4486$B, which 
is an elliptical galaxy at the center of virgo cluster \citep{lau93, lau96}. 
Distribution of stellar peaks in both these galaxies differ from each other: the peaks 
in NGC$4486$B are symmetric w.r.t. the photo-centre in contrast to the off--centre peaks 
in M$31$. Motivated by the work of \citet{tou02}, \citet{ss02} proposed that unstable 
eccentric modes due the presence of counter--rotating streams of matter could be present 
in the nuclei of galaxy M$31$ giving rise to eccentric discs. Counter-rotating streams of 
matter could form possibly due to accretion of stars from in-falling globular clusters. 
Such eccentric discs are thought to be the reason behind observed lopsided multiple--peaked 
brightness distribution around its nuclear black hole, as was proposed by \citet{tre95} for 
M$31$. However, the double-peak stellar distribution in NGC$4486$B, being more symmetrical 
around the photo-centre, is plausibly due to $m = 2$ ($m$ being the azimuthal quantum number) 
unstable modes than $m = 1$ eccentric modes for M$31$ \citep{tre01, ss02, gss12}. 

Unstable eccentric modes are known to exist in self-gravitating counter--rotating streams 
of matter \citep{zh78, ara87, saw88, ms90, pp90, sm94, ljh97, tou02, st10, gss12}. Specifically, 
the counter-rotating discs in the systems discussed above happen to be dominated by the 
influence of a central black hole, thereby making them nearly Keplerian. Moreover, mainly 
comprising stars, these discs are collisionless. Earlier studies of nearly Keplerian 
counter--rotating discs either modelled the disc(s) using a system of rings \citep{tou02}, 
or restricted themselves to softened gravity fluid discs \citep{st10, gss12}---which notably 
supports only $m = 1$ modes---or some similar system. The fluid analysis is inadequate to 
describe systems comprising stellar discs, especially for $m>1$; since using the WKB analysis 
\citet{jt12} showed that slow modes with $m > 1$ exist for nearly Keplerian collisionless discs. 
These form a new class of modes, which have hitherto not been explored in detail.

Recently, \citet{gs16} (Paper~I hereafter) have formulated an integral-eigenvalue equation 
for collisionless self-gravitating disc in the epicyclic approximation using a modified WKB 
formulation. In the present paper we apply this formalism to two coplanar counter-rotating 
discs. The disc profiles considered in this work are the same as in Paper~I. Both discs 
interact with each other only through gravity. We consider the discs in the external potential 
of a central black hole and treat them to be nearly Keplerian systems. Such systems have 
been shown to support {\it Slow modes} that are much slower in comparison to the Keplerian 
flow of the disc. In this work we investigate the properties of these modes as a function of 
mass fraction in retrograde disc and the azimuthal wave number $m$.  

In the next section we introduce the system of unperturbed discs. Thereafter, in \S~\ref{prtrbd_dsc} 
we derive the integral equation for two nearly Keplerian counter-rotating discs. Next, we 
take the slow mode limit to derive the integral eigenvalue equation in \S~\ref{slw_mde}, where 
we show that all $m$ modes are unstable if the mass in retrograde disc is non-zero. We also 
discuss the general properties of the these unstable modes. In \S~\ref{num_mthd}, we discuss 
the details of our numerical method, and in \S-\ref{num_rslts} we discuss the numerical results 
for different values of mass fraction in the retrograde disc. 
We conclude in \S~\ref{cnclsns}.

\section{Unperturbed discs}\label{unprtbd_dsc}
We begin by approximating our discs to be razor thin, i.e., we restrict ourselves 
to $z = 0$ plane, and use polar-coordinates $\mathbf r\equiv (R\,,\phi)$ in the plane of
the discs, with the origin at the location of the central
mass. The unperturbed disc is a superposition of two coplanar 
collisionless counter-rotating discs where the disc particles interact with 
each other gravitationally through Newtonian gravity. Throughout this paper, the 
superscripts `$+$' and `$-$' refer to the prograde and the retrograde discs, respectively.

The unperturbed potential, $\Phi_0(R)$, is the sum of Keplerian potential 
due to the central mass and the self-gravity of both `$\pm$' discs:
\begin{align}
 \Phi_0(R) &\;=\; -\frac{G M}{R} + \Phi_{\rm d}(R)\,,\label{phi}\\
 \Phi_{\rm d}(\bfpar{r}{}{}) &\;=\; -G\int \frac{\sigd{\sss{+}}(\bfpar{r'}{}{}) + \sigd{\sss{-}}(\bfpar{r'}{}{})}
{|\bfpar{r}{}{} - \bfpar{r'}{}{}|} \dmath^2r'\,.\label{phid}
\end{align}
\nin
In this paper we are interested in studying the discs for which 
$\md/M \equiv \varepsilon \ll 1$, where $\md$ is the total mass 
of the disc and $M$ is the central mass. The disc potential $\Phi_{\rm d}$ 
is then on the order $\mathnormal{O}(\varepsilon)$ smaller in comparison to 
the Keplerian potential due to the central mass. Azimuthal and radial frequencies 
($\pm\Omega$ and $\pm\kappa$, respectively) for both `$\pm$' discs are given by
\begin{align}
 \Omega^2(R) & \;=\; \frac{GM}{R^3} + \frac{1}{R}\der{\Phi_{\rm d}}{R}\,,\label{omega}\\
 \kappa^2(R) & \;=\; \frac{GM}{R^3} + \frac{3}{R}\der{\Phi_{\rm d}}{R} + \dder{\Phi_{\rm d}}{R}\,.\label{kappa}
\end{align}
\nin
The precession rate $\pm\pom$ for such near circular orbits is
\begin{align}
 \pom(R) & \;=\; \Omega(R) - \kappa(R)\nonumber\\
& \;=\; -\frac{1}{2\Omega(R)}\left(\frac{2}{R}\der{}{R} + \dder{}{R}\right)\Phi_{\rmpar{d}}(R) 
+ \mathnormal{O}(\varepsilon^2)\,.\label{pom}
\end{align}
\nin 
In the expression for $\pom$ we have retained terms up to linear order in 
$\varepsilon$. For nearly Keplerian discs $\varepsilon \ll 1$; and the slow modes 
in such disc exist due to this small non-zero precession, and 
the complex eigenfrequencies of modes is on the same order as $\pom$.

The disc particles in both prograde and retrograde discs are assumed to be 
in nearly circular orbits, called epicyclic orbits \citep{bt08}. 
We note that $(R', v_R')$ are the same for the `$\pm$' discs, whereas the 
sense of rotation, whether prograde or retrograde, does change the 
expressions for $(\phi', \vtilde')$ for the respective discs. The phase-space 
coordinates of particles for both $\pm$ discs are given by
\begin{align}
R'^{\pm} &\,=\, R + \frac{\gamma\vtilde}{\kappa}\left(1 - \cos(\tau)\right) + \frac{v_R}{\kappa}\sin(\tau)\,,\nonumber\\
\phi'^{\pm} \,=\, \phi \pm \frac{\Omega \tau}{\kappa} &\pm \frac{\gamma\gamma'}{2\kappa}\vtilde\tau \pm 
\frac{\gamma}{R\kappa}\left[\gamma\vtilde\sin(\tau) - v_R\left(1 - \cos(\tau)\right)\right]\,,
\label{epi_approx1}
\end{align}
and
\begin{align}
v_R'^{\pm} &\,=\, v_R\cos(\tau) + \gamma\vtilde\sin(\tau)\,,\nonumber\\
\gamma\vtilde'^{\pm} &\,=\, \pm\left(\gamma\vtilde\cos(\tau) - v_R\sin(\tau)\right)\,,
\label{epi_approx2}
\end{align} 
\nin
where at time $t' = t$, the phase-space coordinates 
$(\bfpar{r'}{}{},\bfpar{v'}{}{}) = (\bfpar{r}{}{},\bfpar{v}{}{})$. 
Also $\vtilde(R) = v_{\phi}(R) - v_c(R)$;\, 
$v_c(R) = R\,\Omega(R)$;\,  $\gamma(R) \;=\; 2\Omega(R)/ \kappa(R)$;\, 
$\tau \,=\, \kappa_g(t' - t)$;\, $\kappa_g \;=\; \kappa(R_g)$ where 
$R_g$ is the mean radius
of the orbit for a given angular momentum; and 
$\gamma'$ is the derivative of $\gamma$ w.r.t. $R$.

The two discs are treated as collisionless and are descried by the Collisionless 
Boltzmann equation (CBE). A solution of CBE in the $z = 0$ plane under the epicyclic 
approximation for axisymmetric stellar discs is given by the Schwarzschild Distribution 
Function (DF) \citep{bt08}, which for $\pm$ disc is given by
\begin{equation}
  f_0^{\pm}(R^{\pm},v_R^{\pm},\vtilde^{\pm}) \;=\; \frac{\gamma \sigd{\pm}(R)}{2\pi\sigma_R^{\pm^2}}\exp
  \left(-\,\frac{v_R^{\pm^2} + \gamma^2\vtilde^{\pm^2}}{2\sigma_R^{\pm^2}}\right)\,,\label{sch2D}
\end{equation}
\nin
where $\sigd{\pm}(R)$ and $\sigma_R^{\pm}(R)$ are the unperturbed surface density profile and 
the radial component of the velocity dispersion, respectively. Note that $\gamma$ is same for 
both `$\pm$' discs.

\section{Perturbed disc: Integral equation for two counter-rotating discs}\label{prtrbd_dsc}
We wish to study the evolution of the discs described above in the linear perturbation 
regime. We begin by perturbing the initial DF's, $f_0^{\pm}$, such that 
\begin{equation}
f^{\pm}(R,\phi,v_R,{\tilde{v}}_{\phi},t) = f_0^{\pm}(R,v_R,{\tilde{v}}_{\phi}) + f_1^{\pm}(R,\phi,v_R,{\tilde{v}}_{\phi},t)\,. 
\label{f_per}
\end{equation}
\nin 
The perturbation $f_1^{\pm}$ are assumed to be $\sim \varepsilon f_0$ and 
henceforth we shall retain terms only up to linear order in the perturbed 
quantities. Volume integral of $f_1^{\pm}$ over velocity space gives the 
perturbed surface density $\siga{1}{\pm}$, i.e. ,
\begin{equation}
\siga{1}{\pm}(R,\phi,t) = \int f_1^{\pm}(R,\phi,v_R,{\tilde{v}}_{\phi},t) \dmath^2\bfpar{v}{}{}\,,\label{den_per}
\end{equation}
where $\dmath^2\bfpar{v}{}{} = \dmath v_R\dmath {\tilde{v}}_{\phi}$. Perturbations in 
the surface density gives rise to perturbed potential, which can be calculated using 
the Poisson integral (see Paper~I). The total gravitational potential at any $(R,\phi,t)$  
is a linear sum of the potential due to the prograde and the retrograde discs. We linearise 
the CBE for both `$\pm$' discs to get 
\begin{equation}
\der{f_1^{\pm}}{t} = - \left[f_0^{\pm}, \Phi_1\right]\,,\label{pcbe1}
\end{equation}
where the time derivative on the left hand side is taken along the 
unperturbed orbit, and the bracket $[*,*]$ on the right hand side 
is the Poisson Bracket. We solve these equation for the prograde and 
the retrograde discs separately for epicyclic orbits as in Paper~I to get
\begin{align}
\siga{a}{\pm}(R) =& \,\frac{2G\siga{d}{\pm}}{R^{5/2}\kappa^2}\sum_{n = 1}^{\infty}\left(\frac{n^2}{n^2 - s^{\pm^2}}\right)
\intgl\frac{\dmath\alpha}{2\pi}\,N(\alpha,m)A_m(\alpha)\,\rme^{\rmi\alpha q}\,B_n(\alpha,\chi^{\pm})\,,
\label{sigdt3}
\end{align}
\nin
where $q = \ln R$, $s^{\pm} = (\omega \mp m\Omega)/(\pm\kappa$), 
$\chi =\sigma_R^{\pm^2}\alpha^2/R^2\kappa^2$, $\omega$ is the temporal 
eigenfrequency, and 
\begin{align}
A_m(\alpha) &\,=\,\intgl\dmath q'\,R'^{3/2}\left[\siga{a}{+}(R') + \siga{a}{-}(R')\right]\rme^{-\rmi\alpha q'}\,,
\label{am}\\
N(\alpha,m) &\, = \,\pi\frac{\displaystyle{\Gamma\left(\frac{m}{2}+
\frac{1}{4}+\frac{\rmi\alpha}{2}\right)
\Gamma\left(\frac{m}{2}+\frac{1}{4}-
\frac{\rmi\alpha}{2}\right)}}
{\displaystyle{\Gamma\left(\frac{m}{2}+\frac{3}{4}+\frac{\rmi\alpha}{2}\right)
\Gamma\left(\frac{m}{2}+\frac{3}{4}-\frac{\rmi\alpha}{2}\right)}}\,,\label{NM_alpha}\\
B_n(\alpha,\chi^{\pm}&) \,=\, \frac{\alpha^2}{\chi^{\pm}}\,\,\rme^{-\chi^{\pm}}\,I_n(\chi^{\pm})\,.
\label{BN}
\end{align}
For mathematical details we refer the reader to Paper~I. Note that $N(\alpha, m)$ 
and $B_n(\alpha, \chi^{\pm})$ are real and even functions of $m$ and $\alpha$ both. 
$I_n(x)$ is the modified Bessel's function of integer order $n$ and $\Gamma(x)$ is 
the Gamma function for a complex argument $x$. From eqn.~(\ref{sigdt3}) it can be 
easily verified that the equations are symmetric under the simultaneous transformations 
$\{$`$+$', $\omega\}$ $\to$ $\{$`$-$', $-\omega\}$, which simply interchanges the meaning 
of prograde and retrograde discs. 

\section{Integral eigenvalue equation for Slow modes}\label{slw_mde}
The presence of an infinite series in the integral equation eqn.~(\ref{sigdt3}), makes 
it difficult to solve it in its present form. As explained in $\S~\ref{unprtbd_dsc}$, 
the presence of a small precession in nearly Keplerian discs allows for the existence 
of slow modes in the discs for which $\Omega \sim \kappa$ and $\pom \sim \mathnormal{O}(\varepsilon)$. 
We make an ansatz that the eigenfrequencies $\omega$ is on the same order as $\pom$, i.e., 
$\omega \sim \mathnormal{O}(\varepsilon) \ll 1$. Using this we find that to leading order 
$s^{\pm} \simeq \mp m$, $\gamma \simeq 2$. Since the infinite summation over $n$ in 
eqn.~(\ref{sigdt3}) contains terms like $n^2/(n^2 - s^{\pm^2})$, the dominant term in 
the series is the one for $n = m$. Also retaining terms up to linear order in $\varepsilon$
\begin{equation}
m^2 - s^{\pm^2} = \frac{2m(\omega \mp m\pom)}{\pm\kappa}\,.
\end{equation}

\nin
We make further simplification by assuming that the velocity 
dispersions in both `$\pm$' discs are equal: $\sigma_R^+ = \sigma_R^- 
= \sigma_R$. This implies $\chi^+ = \chi^- = \chi$.
For convenience we write $\sigd{-} = \eta(R)\sigd{}$ and $\sigd{+} = (1 - \eta(R))
\sigd{}$. $\eta(R)$ is the local mass fraction in the unperturbed retrograde 
disc, $\sigd{}(R) = \sigd{+}(R) + \sigd{-}(R)$ and by definition, 
$0 \le \eta(R) \le 1$. Using all these simplifications 
in eqn.~(\ref{sigdt3}) we get;
\begin{align}
\siga{a}{+}(R) \,=& \,\frac{mG(1 - \eta(R))\siga{d}{}}{R^{5/2}\kappa(\omega - m\pom)}
\intgl\frac{\dmath\alpha}{2\pi}\,\rme^{\rmi\alpha q}\,N(\alpha,m)A_m(\alpha)\,B_m(\alpha,\chi)\,,\label{siga+}\\
\siga{a}{-}(R) \,=& \,\frac{-mG\eta(R)\siga{d}{}}{R^{5/2}\kappa(\omega + m\pom)}
\intgl\frac{\dmath\alpha}{2\pi}\,\rme^{\rmi\alpha q}\,N(\alpha,m)A_m(\alpha)\,B_m(\alpha,\chi)\,.
\label{siga-}
\end{align}
\nin
Above two equations can be used to get a relation between `$\pm$' perturbations, 
which is;
\begin{equation}
\eta(R)\left(\omega - m\pom\right)\siga{a}{+} \,=\, -\left(1 - \eta(R)\right)\left(\omega + m\pom\right)\siga{a}{-}.\label{pmrel}
\end{equation}
We can in principle use this to derive a single equation in either of $\siga{a}{\pm}$, 
solve it, and use the above relation to get the other of $\siga{a}{\pm}$. However, 
the resultant single integral equation is complicated since `$\omega$' occurs inside 
the integral over $R'$. Solving such equation numerically is difficult. Therefore, we adopt 
a different route to solve this system of equations in a later section, however, we shall 
now use this relation to study the general properties of slow modes in this system. 

Using eqn.~(\ref{am}) for $A_m(\alpha)$ in eqn.~(\ref{siga+}) \& (\ref{siga-}) we get
\begin{align}
\mathcal{S^+}(R) &\,=\,
\frac{2m(1 - \eta)}{(\omega - m\pom)}\intgl\dmath q'\,\mathcal{C}(R)\mathcal{C}(R')K_m(\chi, q - q')\,\left[\mathcal{S^+}(R')
+ \mathcal{S^-}(R')\right],\label{siga1+}\\
\mathcal{S^-}(R) &\,=\,
\frac{-2m\eta}{(\omega + m\pom)}\intgl\dmath q'\,\mathcal{C}(R)\mathcal{C}(R')K_m(\chi, q - q')\,\left[\mathcal{S^+}(R')
+ \mathcal{S^-}(R')\right]\label{siga1-}.
\end{align}
where
\begin{align}
K_m(\chi, q) &\,=\, \intglwr{0} \frac{\dmath\alpha}{2\pi}\,\cos(\alpha q) N(\alpha, m) B_m(\alpha,\chi),\label{km}\\
\mathcal{C}(R) \,=\,& \sqrt{\frac{G\sigd{}(R)}{R\kappa(R)}},\,\,\,\,\,\,
{\text{and}}\,\,\,\,\,\,
\mathcal{S^{\pm}}(R) \,=\, \frac{R^{3/2}\siga{a}{\pm}(R)}{\mathcal{C}(R)}\label{c&spm}.
\end{align}
\nin
Note that in eqn.~(\ref{km}) the integral over $\alpha$ is from zero to infinity rather then from 
$-\infty$ to $\infty$. We could make this simplification since both $N(\alpha, m)$ and $B_m(\alpha, \chi)$ 
are even functions of $\alpha$. Adding the equations for $\mathcal{S^{\pm}}(R)$ and defining 
$\mathcal{S}(R) = \mathcal{S^+}(R) + \mathcal{S^-}(R)$ we can obtain a single integral equation in 
$\mathcal{S}(R)$
\begin{align}
\left(\frac{\omega^2 - m^2\pom^2}{\omega(1 - 2\eta(R)) + m\pom}\right)\mathcal{S}(R) &\,=\,
2m\intgl\dmath q'\,\,\left[\mathcal{C}(R)\mathcal{C}(R')K_m(\chi, q - q')\right]\,\,\mathcal{S}(R').
\label{int_eqn}
\end{align}

\nin
We can solve the above integral equation for the unknown $\mathcal{S}(R)$ and the eigenvalue 
`$\omega$'. Then we use relation (\ref{pmrel}) along with the definition of $\mathcal{S}(R)$ 
to recover $\mathcal{S}^{\pm}(R)$;
\begin{align}
\mathcal{S^+}(R) = (1 - \eta)\frac{\omega + m\pom}{(1 - 2\eta)\omega + m\pom}\mathcal{S}(R), \,\,\,\,\,\,\,
\mathcal{S^+}(R) = - \eta\frac{\omega - m\pom}{(1 - 2\eta)\omega + m\pom}\mathcal{S}(R).\label{pmrel1}
\end{align}
\nin
Before going any further to discuss the nature of solution for the above integral eigenvalue 
equation we shall first make certain assumptions regarding the velocity dispersion 
profile and surface density profiles for both `$\pm$' that we shall use in this paper 
to solve the integral eigenvalue problem formulated above.

\nin
{\it Velocity Dispersion:} As used in Paper~I, and also suggested by \citet{jt12} as a 
reasonable profile for velocity dispersion, we take 
$\sigma_R = \sigma R\kappa(R) \simeq \sigma R\Omega = \sigma v_c(R)$, where $\sigma < 1$ 
in order to satisfy the epicyclic condition and is a constant. The second equality here is 
due to near Keplerian nature of orbits for slow modes. This profile simplifies the integral 
equation immensely as $\chi$ becomes a constant. Using this also allows us comparison of 
our work with earlier works by \citet{tre01, jt12}. 

\nin
{\it Surface density:} Henceforth, we shall assume that $\eta$ is a constant. This would 
imply that both `$\pm$' discs have similar radial profiles for surface density. The case of 
single disc in Paper~I corresponds 
to $\eta = 0$ (or $1$) with $\omega$ ($-\omega$) giving the corresponding 
eigenvalues. We shall numerically solve the integral equation 
for the following two surface density profiles:
\begin{itemize}
 \item {\it Kuzmin Disc:} First, we use the Kuzmin disc profile, 
 which has a centrally concentrated disc profile given by
 \begin{align}
 \sigd{\rm Kz}(R) \,=\, & \frac{a\md}{2\pi(a^2 + R^2)^{3/2}}\,,\label{sig_kzn}\\
 \pom^{\rm Kz}(R) \,=\, & -\frac{3G\md a^2}{2\Omega(R)(a^2 + R^2)^{5/2}}\,\label{pom_kzn},	
 \end{align}
 \nin
 where $a$ is the concentration parameter. 
 \item {\it JT annular disc:} The second is an annular disc introduced 
 by \citet{jt12}
 \begin{align}
  \sigd{\rm JT}(R) \,=\, &  \frac{3\md b R^2}{4\pi(b^2 + R^2)^{5/2}}\,,\label{sig_tmre}\\
  \pom^{\rm JT}(R) \,=\, & \frac{3G\md b^2 (b^2 - 4R^2)}{4\Omega(R)(b^2 + R^2)^{7/2}}\,.\label{pom_tmre}  
 \end{align}
 \end{itemize}
 
 \nin
Here $b$ is a length scale.
Both these profiles are physically quite different and hence form very 
good test cases to be explored. Also same profiles have been used in Paper~I, 
and also by other authors \citep{tre01,jt12,gss12} to study similar problems.

\subsection{Dispersion relation and stability analysis}\label{stability} 
We pause here to derive the 
dispersion relation for counter--rotating discs and analyze the stability of 
modes. Stationary phase approximation can be used to solve the integrals over 
$q'$ and $\alpha$ in eqns.~(\ref{sigdt3}) - (\ref{am}) under the limit 
$\alpha \gg m$. In Paper~I authors have used this method to take the local limit 
of the integral equation derived for single disc and show that their equation 
reduces to the well known WKB dispersion relation of    
\citet{tmre64}. 

Using exactly the same approximations, and combining eqn.~(\ref{sigdt3}) \& (\ref{am}) we get
\begin{align}
\siga{a}{\pm}(R) =& \frac{2\pi G\sigd{\pm}(R)|k|}{\kappa^2}\sum_{n = 1}^{\infty}\left(\frac{n^2}{n^2 - s^{\pm^2}}\right)
\frac{2}{\chi}\rme^{-\chi}I_n(\chi)\siga{a}{}(R)\,,\label{sigdt6}
\end{align}
\nin
where $\siga{a}{} = \siga{a}{+} + \siga{a}{-}$. We do not give details of algebra here, 
the interested readers can refer to Appendix~A of Paper~I for more details. Adding and rearranging the terms we get
\begin{equation}
2\pi G|k| \left(\frac{\mathcal{F}^{\sss{+}}}{D_m^{\sss{+}}}\sigd{\sss{+}} + \frac{{\mathcal{F^{\sss{-}}}}}
{D_m^{\sss{-}}}\sigd{\sss{-}}\right) = 1,\label{disp_gen}
\end{equation}
where
\begin{align}
D_{m}^{\sss{\pm}}  \,=\,\, & \kappa^2 - (\omega \mp m\Omega)^2,
\label{dmpm}\\
 \mathcal{F}^{\sss{\pm}}(s^{\pm},\chi) \,=\,\, & \frac{2}{\chi}\,(1 - s^{\pm^2})\, e^{-\chi} 
 \,\sum_{n=1}^{\infty}\frac{I_{n}(\chi)}{1 - s^{\pm^2}/n^2}\,.
\label{Fpm}
\end{align}

For $m = 0$, the dispersion relation \eqref{disp_gen} reduces to the well-known 
relation due to \citet{tmre64}. This implies that  
the counter--rotating discs are stable to axisymmetric perturbations if 
$Q \equiv \sigma_R\kappa/3.36G\sigd{} > 1$. Since in this paper we are interested 
in studying the properties of eigenmodes in near Keplerian discs, we now reduce 
the dispersion relation in eqn.~(\ref{disp_gen}) to specialize to slow modes with 
$\omega \sim \mathnormal{O}(\varepsilon)$ for $m \ge 1$. 
In this case $s^2 \to m^2$, and the dominant term in the summation in expression for 
$\mathcal{F}^{\sss{\pm}}$ corresponds to
$n = m$. Applying this and keeping terms up to linear order in small 
quantity $\varepsilon$ we get
\begin{align}
\frac{\mathcal{F}^{\sss{\pm}}}{D_{m}^{\sss{\pm}}} &= \frac{\pm m \mathcal{F}_{m}(\chi)}
{2\Omega(\omega \mp m\pom)},\nonumber\\
\mathcal{F}_{m}(\chi) &= \frac{2}{\chi}{\rmpar{e}}^{-\chi}I_m(\chi)\,.
\end{align}
\nin 
Using this in equation~(\ref{disp_gen}) and using $\eta =
\sigd{\sss{-}}(R)/\sigd{}(R)$ 
to be the mass fractions in retrograde disc, we get
\begin{equation}
 \omega^2 + B_m\omega + C_m = 0\,,\label{disp_gen_slow}
\end{equation}
where
\begin{align}
 B_m &= \frac{-\pi mG|k|\sigd{}\mathcal{F}_{m}}{\Omega}\left(1 - 2\eta\right),\nonumber\\
C_m &= -m^2\pom^2 -\frac{\pi m^2G|k|\sigd{}\pom\mathcal{F}_{m}}{\Omega}.\label{def_b&c}
\end{align}

The above equation is quadratic in $\omega$ and its discriminant 
$D$ is
\begin{equation}
D = m^2\left[\nu^2 (1 - 2\eta)^2\mathcal{F}_m^2 + 4\pom \nu \mathcal{F}_m + 4\pom^2\right]\,,\label{discriminant}
\end{equation}  
\nin 
where $\nu = \pi G|k|\sigd{}/\Omega$. 
Modes are unstable if $D < 0$ and stable otherwise. 
For the case of single disc ($\eta = 0$), the discriminant 
$D = m^2\left(\nu\mathcal{F}_m + 2\pom\right)^2 \ge 0$, which 
implies the modes are all stable, as we know already from 
\citet{jt12} and Paper~I. For non-zero counter--rotation $D > 0$ 
when $\pom > 0$. This agrees with the previous findings 
of \citet{st10, gss12}. However $\pom < 0$ for the mass 
precession in most realistic discs and also is the case 
for the test surface density profiles chosen for the present paper. Here 
we briefly note some general conclusions for $\pom < 0$ discs:

\begin{enumerate}
\item For equal counter--rotation $\eta = 1/2$, and equation~(\ref{disp_gen_slow}) 
says that $\omega^2 = m^2\left(\pom^2 + \nu\pom\mathcal{F}_m\right)$ should be a 
real quantity. Modes are stable and oscillatory if $|\pom| > \nu\mathcal{F}_m$, and 
purely growing/damping otherwise. Thus the stability condition is
\begin{equation}
\frac{\sigma_R}{v_c} > \left|\frac{\sigma_{R0}}{R\pom}\right|\,,
\end{equation}
here $v_c$ is the circular velocity and we have defined $\sigma_{R0} = 
\pi G\sigd{}/\Omega$. In writing the above we have 
used the fact that $2 \rme^{-\chi}I_m(\chi)/\sqrt{\chi} < 1$ for all 
values of $\chi$ and $m$.

\item Defining $\mathcal{H} = \nu\mathcal{F}_m$, solution for the relation $D = 0$
  is given by
\begin{equation}
 \mathcal{H}_{\sss{\pm}} = 2|\pom|\left[\frac{1 \,\pm\, 2\sqrt{\eta(1 - 
\eta)}}{(1 - 2\eta)^2}\right].
\end{equation}

\nin 
Note that $\mathcal{H}$ is always positive. It is straightforward 
to determine that for $0 < \mathcal{H}_{\sss{-}} <  \mathcal{H} < 
\mathcal{H}_{\sss{+}}$, we have $D < 0$. Hence the system is stable 
if $\mathcal{H}_{\sss{-}} > \mathcal{H}_{\rm max}$, where $\mathcal{H}_{\rm max}$ 
is the maximum value of $\mathcal{H}$ in the disc which is calculated below. For 
Keplerian discs $\mathcal{H}$ can be written as
\begin{align}
 \mathcal{H} = & \frac{\pi G\sigd{}\kappa}{\Omega\sigma_R}\frac{2}{\sqrt{\chi}}
\rme^{-\chi}I_m(\chi)\nonumber\\
< & \frac{\Omega\sigma_{R0}}{\sigma_R} 
\equiv  \mathcal{H}_{\rm max}\,.
\label{smax}
\end{align}
Using all this, the condition $\mathcal{H}_{\sss{-}} > \mathcal{H}_{\rm max}$ simplifies to
\begin{equation}
\frac{\sigma_R}{v_c} > \left|\frac{\sigma_{R0}}{R\pom}\right|\left[\frac{(1 - 2\eta)^2}
{2 - 4\sqrt{\eta(1 - \eta)}}\right]\,,
\label{stbl_mne0}
\end{equation}
\nin 
Terms on the RHS are
of $\mathnormal{O}(1)$ whereas for the unperturbed distribution function 
assumed in the present formulation, that is Schwarzschild distribution 
function, it is 
assumed that $\sigma_R/v_c \ll 1$. So discs are largely unstable. 
Note that in deriving Eq.~(\ref{stbl_mne0}) we have not used 
any constraint on the value of $\eta$ and it is applicable for 
all the values of $\eta \ne 0$.
For example, when $\eta \to 1/2$, the term in $[\,\dots\,]$ goes to 
unity and the condition (\ref{stbl_mne0}) reduces to the one 
derived for $\eta = 1/2$ above.
\end{enumerate}

We next aim to solve the integral eigenvalue equation derived in Eq.~(\ref{int_eqn}) 
numerically to study the properties of eigenmodes, 
but before going into the details of the numerical method adopted and numerical 
solutions, we shall give some general conclusions regarding the nature of 
the eigenmodes which can be drawn from the integral eigenvalue 
problem:
\begin{itemize}
 \item With the choice of $\sigma_R$ we have made, $\chi$ is a constant.
 Using this it can be easily verified that the kernel of the integral in 
 eqn.~(\ref{int_eqn}) is real and symmetric in $(R, R')$, which implies 
 that either the eigenvalues are real, or exist in complex conjugate pair for all values of $\eta$.
 
We next consider two special values of $\eta$:
 \item {$\mathbf{\eta = 0\,\,\text{or}\,\,1:}$} This case corresponds to the case 
 of no counter-rotation. Using $\eta = 0$ or $1$ in eqn.~(\ref{int_eqn}) the 
 l.h.s. reduces to $(\pm\omega - m\pom)\mathcal{S}(R)$, and hence we get $\omega$ 
 is always real and the eigenfunctions can be taken real. Thus, the slow modes 
 are stable and oscillatory in time for a single disc.
 \item {$\mathbf{\eta = 1/2:}$} This value of $\eta$ corresponds to the case of 
 equal mass in both prograde and retrograde discs and the net angular momentum in the 
 disc is zero. Substituting the value of $\eta$ 
 in eqn.~(\ref{int_eqn}) we get $\omega^2$ is always real implying that the slow 
 modes are either stable and oscillatory in time or purely growing/damping modes. 
 When modes are stable the eigenfunctions $\mathcal{S^{\pm}}(R)$ can be taken to be 
 real. On the other hand when `$\omega$' is purely imaginary, $\mathcal{S}(R)$ can 
 be taken as a real function multiplied by an arbitrary constant. There are two special 
 cases: (i) If $\mathcal{S}(R)$ is purely real, then the eqn~(\ref{pmrel1}) can be 
 used to verify that $\mathcal{S^{\pm}}(R)$ are complex conjugate of each other. 
 (ii) If $\mathcal{S}(R)$ is a purely imaginary number, then the same eqn.~(\ref{pmrel1}) 
 can be used to see that $\mathcal{S^+}(R)$ is equal to negative of complex conjugate of 
 $\mathcal{S^-}(R)$ and vica-versa.
\end{itemize}
The above conclusions are consistent with the earlier work by \citet{tre01, st10, gss12, jt12}. 
The only difference in the results presented above and the work of \citet{gss12} is for the case of 
$\eta = 1/2$ and $\omega$ purely imaginary. The relation between `$\pm$' perturbations when 
$\mathcal{S}(R)$ is purely real or imaginary are with opposite signs in both. The reason for this 
is that in both the work the `$\pm$' perturbations are linearly combined to get one single integral 
equation with opposite signs and hence the difference. Thus we can attribute this change just 
to the difference in way things are defined and nothing changes qualitatively. 
Later in this paper we shall 
do a detailed quantitative comparison with the work of these authors and comment more on the 
pros and cons of different approaches. To make further progress, in next sections we numerically 
solve eigenvalue problem for given value of $\eta$ and $m$.

\section{Numerical method}\label{num_mthd}

In this section we give the method used to solve the integral eigenvalue problem numerically. 
The integral equation written for $\mathcal{S}(R)$ given in eqn.~(\ref{int_eqn}) can be 
solved to give the value of $\mathcal{S}(R)$ and $\omega$ and then we can use relation 
given in Eqn.~(\ref{pmrel1}) to get $\mathcal{S^{\pm}}(R)$. But this route is not 
very convenient for values of $\eta$ other then `$0$ \& $1/2$' since for other values 
$\eta$ we will get a quadratic eigenvalue problem, which is computationally more expensive 
to solve. It is better to solve the set of coupled integral equation given by 
eqn.~(\ref{siga1+})--(\ref{km}) for $\mathcal{S^{\pm}}(R)$ and $\omega$. As we shall 
see below these can be reduced to a simple eigenvalue problem. 

The first step is to convert the equation to a dimensionless form. We shall normalize the 
radius $R$ of the disc using a length scale $L$. This length scale is present in both the 
surface density profiles we wish to use, `$a$' for Kuzmin disc and `$b$' in case of JT annular 
disc. Other physical quantities can be made dimensionless by defining the characteristic 
surface density by $\md/L^2$ and characteristic orbital frequency by $\Omega^* = \sqrt{GM/L^3}$. 
The result of using these is the rescaling of eigenvalue $\omega$ by $(\Omega^* L^3/G\md)$, 
making it dimensionless. All the notations used earlier will stand for dimensionless quantities hereafter.

Method adopted to solve the integral equation is exactly the same as that used in Paper~I for 
a single disc; here we generalize it to a coupled system of counter-rotating discs. We first rewrite eqn.~(\ref{siga1+}) 
\& (\ref{siga1-}) in the following form
\begin{align}
(\omega - m\pom)\mathcal{S^+}(R) &\,=\,
m(1 - \eta)\intgl\dmath q'\,\,\mathcal{G}_m(\chi, R, R')\,\left[\mathcal{S^+}(R')
+ \mathcal{S^-}(R')\right],\label{siga2+}\\
(\omega + m\pom)\mathcal{S^-}(R) &\,=\,
-m\eta\intgl\dmath q'\,\,\mathcal{G}_m(\chi, R, R')\,\left[\mathcal{S^+}(R')
+ \mathcal{S^-}(R')\right]\label{siga2-}.
\end{align}
where
\begin{align}
 \mathcal{G}_m(\chi, q, q') \,=\, & 2\,\mathcal{C}(R)\,\mathcal{C}(R')\,K_m(\chi, q - q'), 
\end{align}
\nin
Note that $q = \ln R$. First we need to calculate $\mathcal{G}_m(\chi, q, q')$, which involves 
the calculation of functions $\mathcal{C}(R)$ and $K_m(\chi, q - q')$. First one is a simple 
algebraic function when substituted for $\sigd{}(R)$ and $\kappa(R)$. Note that with the 
functional form of velocity dispersion we are using, `$\chi$' is a function of $\sigma$ and 
$\alpha$ only. Hence
\begin{align}
K_m(\chi, q) \,\equiv\, K_m(\sigma, q) \,=\, \intglwr{0} \frac{\dmath\alpha}{2\pi}\,\cos(\alpha q) N(\alpha, m) B_m(\alpha,\chi)\,,
\end{align}
\nin
which is the same as in Paper~I, and we adopt the same method as used there. We do not give the 
details here and refer the interested readers to Paper~I for details. We tabulate $K_m$ as 
a function of $q$ for given values of $m$ and $\sigma$ for $q$ ranging from $[-14,14]$. 

Next we discretise the integral eqns.~(\ref{siga2+}) \& (\ref{siga2-}). The chosen 
range for $-6 \le q \,({\text{and}\, q')} \le 6$ is divided into a grid of 
$n_q$ points using Gaussian quadrature rule. We use a finite range of $q$ and $q'$ 
to avoid numerical singularities at $q\,\,(\text{or}\,\,q') \to -\infty$. Also towards the other 
end, that is for larger radii, the surface density in the disc is very low due to which 
the contribution of integrand towards the tail is negligible. The integral over $q'$ in 
eqns.~(\ref{siga2+}) \& (\ref{siga2-}) are then discretized using
\begin{equation}
\int_{-\infty}^{\infty} {\mathrm{d}q'}\,
\mathcal{G}_m(\sigma,q_i,q')\,\mathcal{S}(q') \quad\longrightarrow\quad \sum_{j=1}^{n_q}w_{q_j}\,\mathcal{G}_m(\sigma,q_i,q_j)\,
\mathcal{S}(q_j)\,,
\label{disct_schm}
\end{equation}

\nin
where $w_{q_j}$ are the weights chosen from the Gaussian quadrature rule and 
as defined earlier $\mathcal{S}(q')  = \mathcal{S^+}(q') + \mathcal{S^-}(q')$. 
Note that in the argument of $\mathcal{G}_m$, we have replaced $\chi$ with 
$\sigma$ as was done for $K_m$ earlier. Then this discretized integral can 
be used to write the matrix eigenvalue problem as

\begin{equation}
\bf{A}\,\zeta \;=\; \omega\,\zeta\,,\label{eigeqn} 
\end{equation}
\nin
where 
\begin{equation}
\bf{A} \;=\; \left[\begin{array}{cc}
                         m(1-\eta)w_j {\mathcal G}_{ij} + m\pom_j\delta_{ij} \quad&\quad m(1-\eta)w_j{\mathcal G}_{ij}\\[1em]
                        -m\eta w_j{\mathcal K}_{ij} \quad&\quad -m\eta w_j {\mathcal G}_{ij} - m\pom_j\delta_{ij}
                        \end{array}\right]\,,\quad \mbox{and}\quad
\bf{\zeta} \;=\; \left(\begin{array}{c}
                         \mathcal{S}^{+}_{i}\\[1em] \mathcal{S}^{-}_{i}
                        \end{array}\right) \,.
\end{equation}

\nin
The above matrix $\bf{A}$ is a $2n_q \times 2n_q$ matrix written in the form of 
$2 \times 2$ block matrix, with each block given by a $n_q \times n_q$ matrix. 
In the above representation $i$ \& $j$ are respectively the row and column 
indices of each $n_q \times n_q$ block. The kernel of the original integral equation 
was symmetric, but the use of unequal weights destroys the symmetry. Same problem 
was encountered in Paper~I and the symmetry can be resolved using a transformation given 
in $\S 18.1$ of \citet{prs92}, details of which are also discussed in Paper~I. 
The above matrix eigenvalue problem yields $2n_q$ eigenvalues and eigenvectors. 
The eigenvector is a $2n_q \times 1$ column vector, where the first $n_q$ entries 
give $\mathcal{S}^+$ and the next $n_q$ entries give $\mathcal{S}^-$. Many of 
these $2n_q$ eigenvalues are singular (van Kampen) modes as is also concluded in 
Paper~I and also by other authors earlier \citep{tre01, gss12, jt12}. Next we solve 
the above matrix eigenvalue problem using the linear algebra package LAPACK \citep{lapack} 
to calculate the eigenvalues and eigenvectors and discuss the properties of the eigenspectrum 
and waveforms we get in the section.

\section{Numerical results}\label{num_rslts}
We solve the matrix equation for both the surface density profiles discussed earlier 
for various values of $\eta$. As noted earlier, the equations are symmetric under 
the transformation $(\eta, \omega) \to (1 - \eta, -\omega)$, this is just 
interchanging the meaning of prograde and retrograde orbits. Hence it is sufficient 
to choose values of $\eta$ in the range $0 \le \eta \le 1/2$. 
For each value of $\eta$ and surface density model we solve for $m = 1\,\&\,2$ and 
$\sigma = 0.1, 0.2, 0.3\,\&\,0.4$. We begin with solving for $\eta = 0$ to benchmark 
our numerics. This case corresponds to a single disc whose particles are rotating in 
a prograde sense. All the $2n_q$ eigenvalues we get are real and the spectrum is similar 
to that in Paper~I. Discrete eigenspectrum we get is exactly same as that of Paper~I. 
The extra $n_q$ eigenvalues, which are essentially $\omega = -m\pom$ corresponds to 
singular (van Kampen) modes. We refer the reader to Paper~I for a discussion of these 
singular modes. Below we give some properties of the spectrum for $\eta = 0$;

\begin{enumerate}
 \item Most of the eigenvalues that we get constitute the continuous part of the spectrum 
 corresponding to singular (van Kampen) modes. 
 \item The non-singular (discrete) eigenvalues are prograde, i.e. non-singular values 
 $\omega$ are positive.
 \item For a given value of $m$, the largest eigenfrequency is a decreasing function of $\sigma$.
 \item For a given value of $\sigma$, the largest eigenvalue decreases as we go from $m = 1$ to $2$. 
 Also as compared to $m = 2$ the eigenfunctions are more radially compact for $m = 1$.
 \item With the decreasing value of $\omega$, the wavelength of the oscillations decreases 
 whereas the number of nodes increases. Also larger the value of $\sigma$, more is the 
 compactness of the wavepackets.
 \item All the above properties are same for both the surface density profiles chosen. 
\end{enumerate}
\nin
In the rest of this section we present the results for other values of $\eta$. 

\subsection{Equal counter-rotation}\label{eq_cr}
\begin{figure}
\begin{center}
\includegraphics[angle=0,scale=0.64]{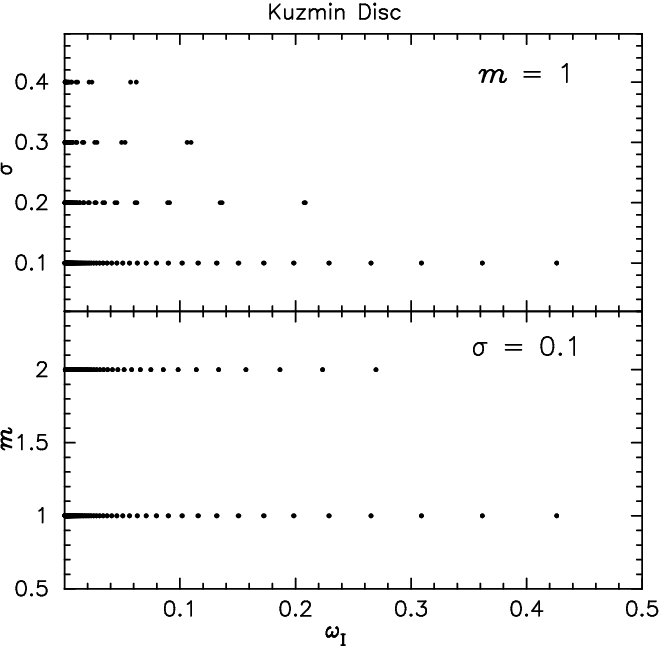}
\includegraphics[angle=0,scale=0.64]{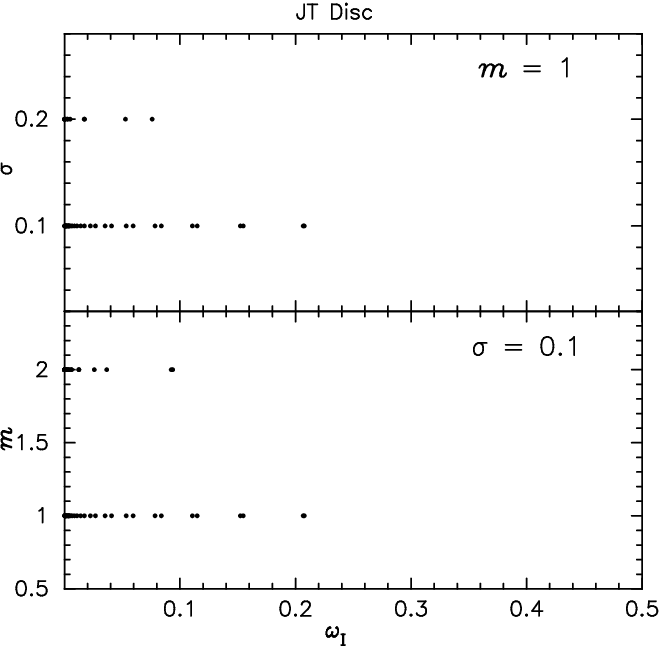}
\caption{Eigenvalue plot for $\eta = 1/2$. Horizontal axis is the growth rate 
whereas the vertical axis is the value of dimensionless velocity dispersion (top 
panel) and $m$ (bottom panel). Plots are labelled for the unperturbed surface 
density profile, $m$ value in top panel and $\sigma$ value in the bottom panel.}
\label{omga_I_vs_sig_m}
\end{center}
\end{figure}

Here we present the results for the case of $\eta = 1/2$. \citet{gss12,st10} have done a similar 
study for a zero pressure softened gravity disc, which supports only $m = 1$ modes. 
As pointed earlier, for $\eta = 1/2$ the eigenvalues are either purely 
real, i.e. oscillatory modes, or purely imaginary, i.e. growing/damping modes. Here we are 
mainly interested in the properties of imaginary eigenvalues. For this case the eigenvalue 
$\omega$ can be written as $\omega = \pm\rmi\omega_{\text{\tiny{I}}}$, 
where $\omega_{\text{\tiny{I}}}$ is the growth rate of the eigenmodes. 
Fig.~\ref{omga_I_vs_sig_m} is the plot of growth rate versus $\sigma$ 
for $m = 1$ (top panel) and $\omega_{\text{\tiny{I}}}$ versus $m$ for 
$\sigma = 0.1$ (bottom panel). Left panel is for Kuzmin disc whereas the 
right one is for JT annular disc. Let $\omega_{\text{\tiny{I}}_{\rm max}}(\sigma,m)$ 
be the maximum value of growth rate for a given $(\sigma, m)$. The general trends 
observed in the spectrum are: (1) For a given value of $m$, 
$\omega_{\text{\tiny{I}}_{\rm max}}$ is a decreasing function 
of $\sigma$. (2) For a given value of $\sigma$, $\omega_{\text{\tiny{I}}_{\rm max}}$ decreases 
as we go from $m = 1$ to $m = 2$. 

The eigenvalues exists as degenerate pairs, which are also present for the case of single 
disc. The eigenvalues are so closely spaced that we can hardly 
distinguish them in fig.~\ref{omga_I_vs_sig_m}. In Table~\ref{table_1} we give values of a few 
such eigenvalues, beginning from the largest value of growth rate for Kuzmin disc for $m = 1$ 
and $\sigma = 0.1, 0.2 \,\,\&\,\, 0.3$. The pairs form due to the existence of leading and 
trailing waves. The separation between the degenerate pair increases as we go 
to higher values of $\sigma$. Also for a given value of $\sigma$, the separation in the 
eigenvalue pair increases as the growth rate decreases. 

\begin{figure}
\begin{center}
\includegraphics[angle=0,scale=0.54]{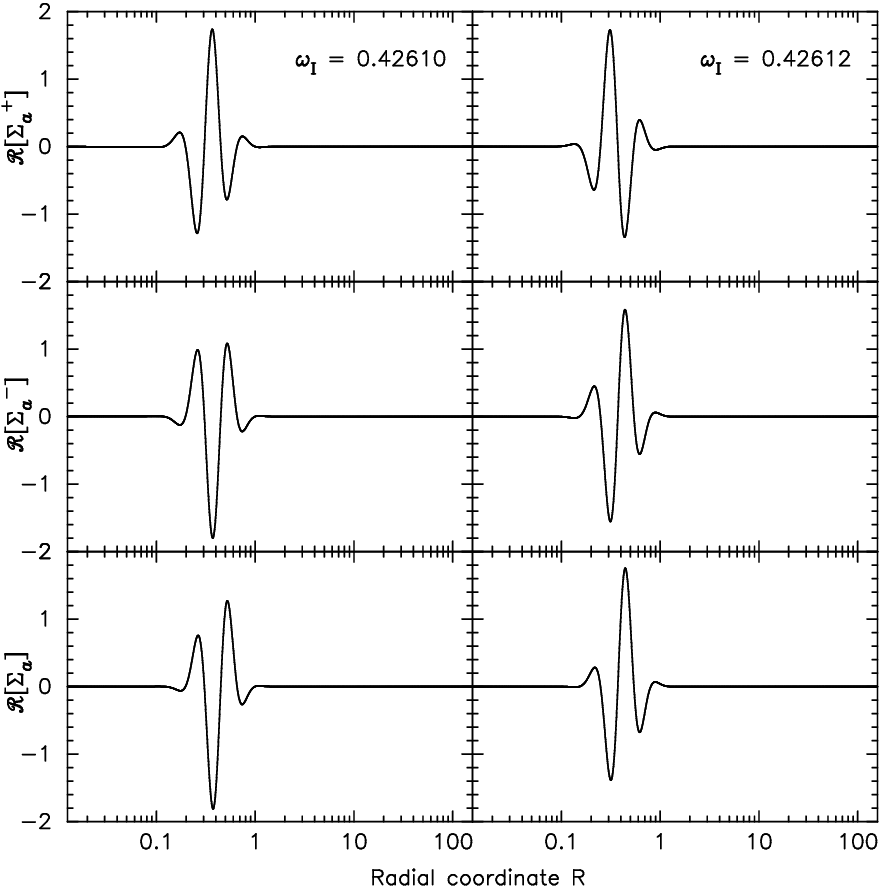}
\includegraphics[angle=0,scale=0.54]{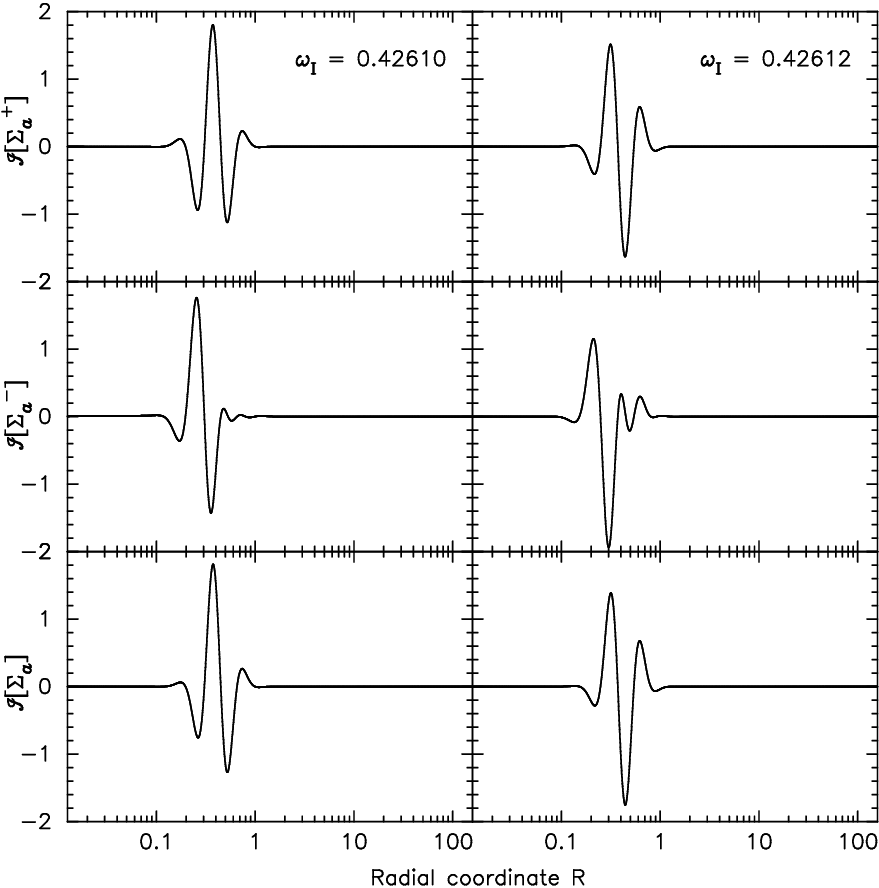}
\caption{This plot displays the radial variation of real (left panel) and imaginary 
(right panel) components of $\siga{a}{+}$, $\siga{a}{-}$ and 
$\siga{a}{} = \siga{a}{+} + \siga{a}{-}$. Unperturbed surface disc profile is 
Kuzmin disc, $m = 1$ and $\sigma = 0.1$. Plot is made for degenerate pair of eigenvalues, 
and the top panel of each column is labelled for the value of $\omega_{\text{\tiny{I}}}$}
\label{eigplot_1}
\end{center}
\end{figure}

\begin{figure}
\begin{center}
\includegraphics[angle=0,scale=0.7]{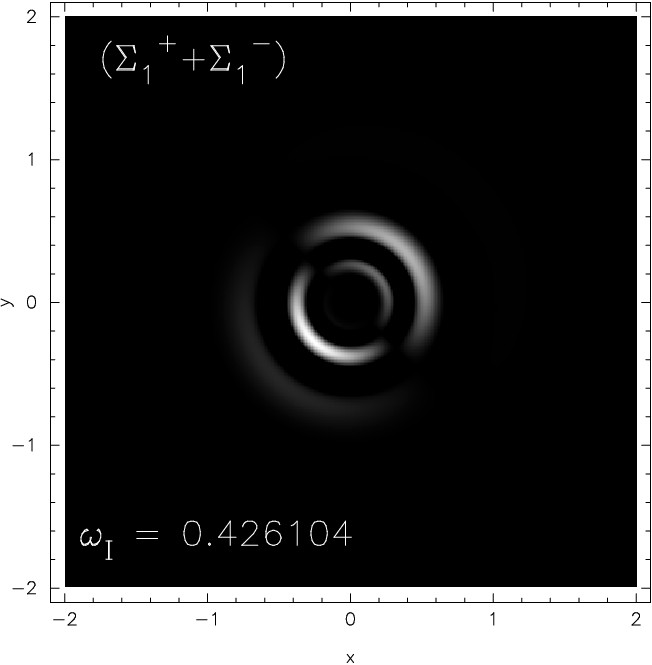}
\includegraphics[angle=0,scale=0.7]{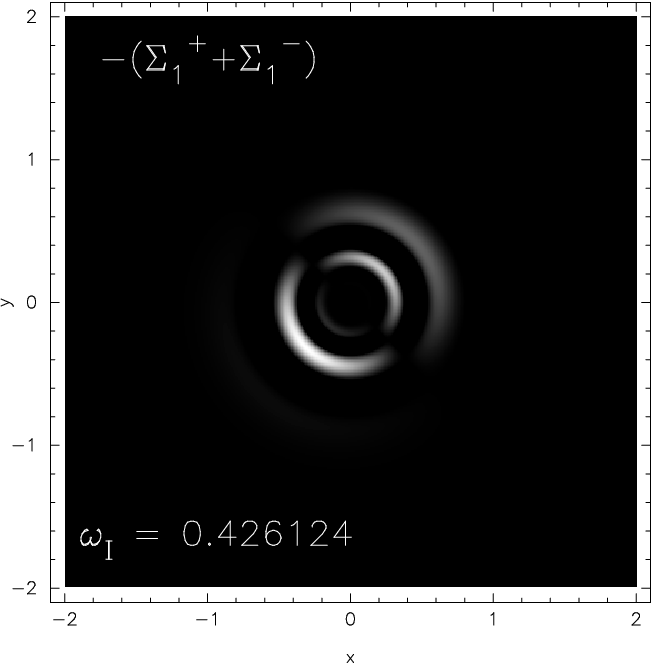}
\caption{Gray-scale image of the positive component of real part of 
total surface density profile in $x-y$ plane at time $t = 0$. White and 
black represent maximum and minimum/zero in the surface density for the 
model parameters same as that of fig.~\ref{eigplot_1}. Leading and trailing 
wave behaviour of the degenerate pairs of eigenvalues can be seen in the 
images. Plotting the negative of $\siga{a}{}$ is motivated from the radial 
profile in the bottom panel of fig.~\ref{eigplot_1}.}
\label{eigplot_2}
\end{center}
\end{figure}

In fig.~\ref{eigplot_1} we give the plot of radial variation of $\siga{a}{\pm}$ and 
$\siga{a}{} = \siga{a}{+} + \siga{a}{-}$ for the first degenerate pair of eigenvalue. 
Left panel is the plot of real part of the eigenfunction whereas right panel is the 
plot of imaginary part for the pair degenerate eigenvalues. Kuzmin disc is used as 
the unperturbed disc profile. $\sigma \,\,\&\,\, m$ values used are $0.1\,\,\&\,\,1$, 
respectively. Panels are labelled for the value of growth rate. The radial variation of 
total surface density, i.e. the bottom panel shows the leading and trailing wave behaviour. 
We illustrate this further in fig.~\ref{eigplot_2}, where we plot a gray scale image of 
the density enhancement for the real part of the total surface density $\siga{1}{+} + 
\siga{1}{-}$ in $x$-$y$ for the same disc parameters as fig.~\ref{eigplot_1}. 
White/black gives the maximum/minimum (or zero) surface density. In the right panel we 
have taken  $-(\siga{1}{+} + \siga{1}{-})$. Since we have restricted ourself to linear 
analysis, the eigenfunctions we get are known only up to a constant multiplicative factor. 
Minus factor is motivated by the inspection of the lower panel of fig.~\ref{eigplot_1}. 
The leading and trailing behaviour of the degenerate pairs of eigenvalues can be clearly 
seen on comparing both panels in fig.~\ref{eigplot_2}.

\begin{figure}
\begin{center}
\includegraphics[angle=0,scale=0.665]{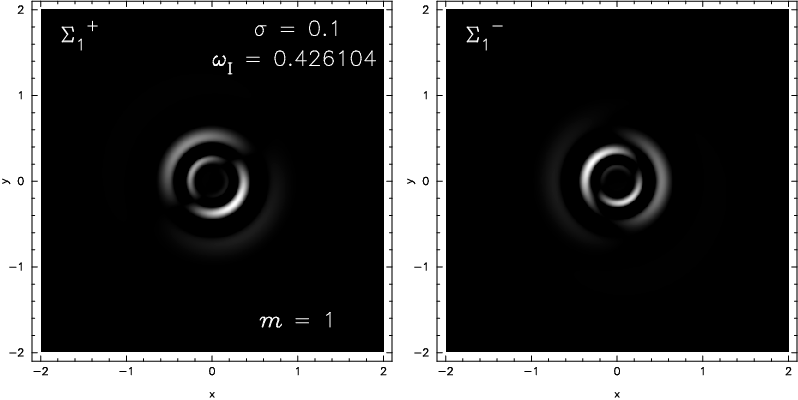}
\includegraphics[angle=0,scale=0.4]{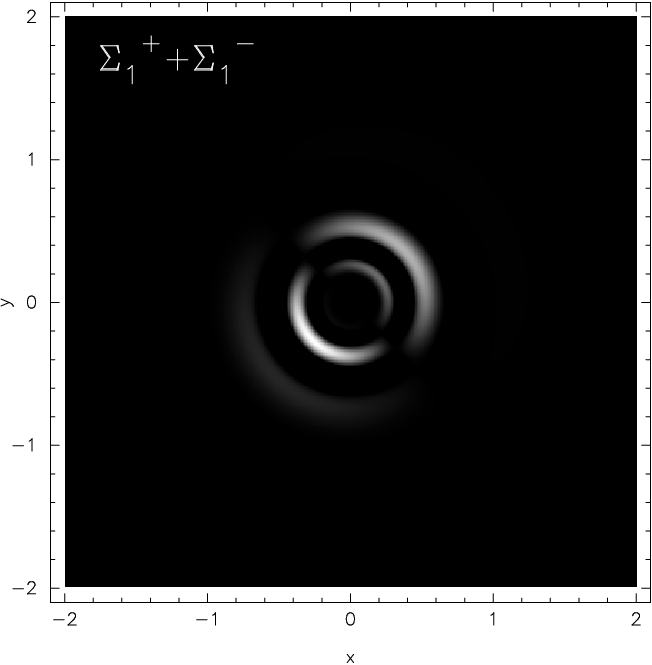}
\includegraphics[angle=0,scale=0.665]{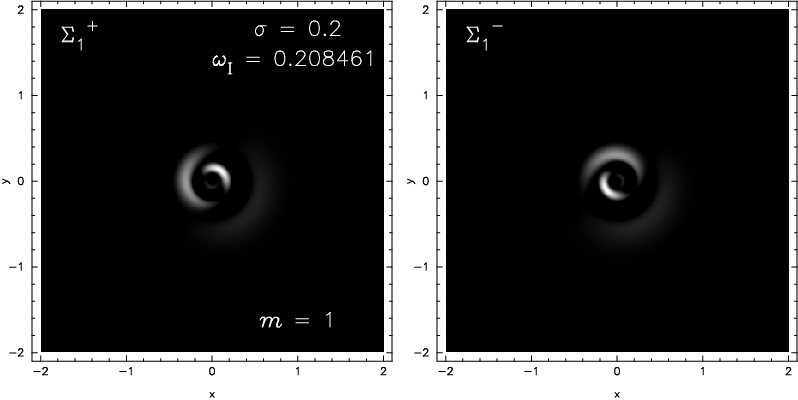}
\includegraphics[angle=0,scale=0.4]{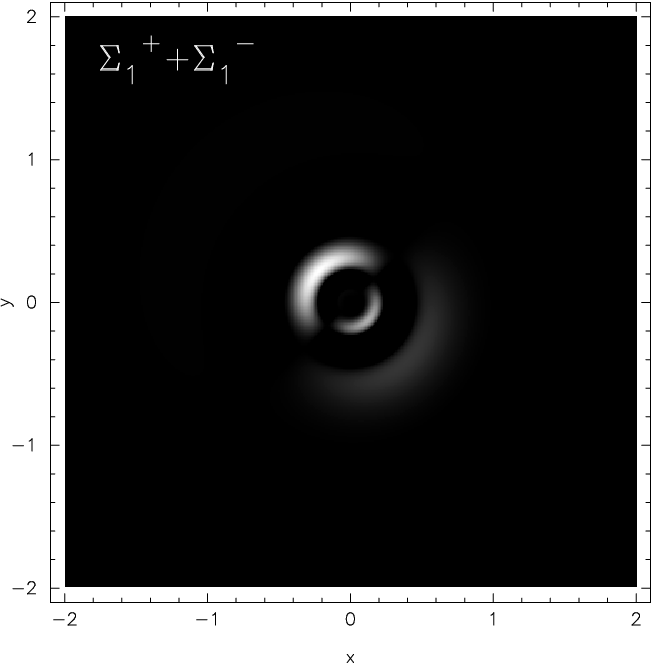}
\includegraphics[angle=0,scale=0.665]{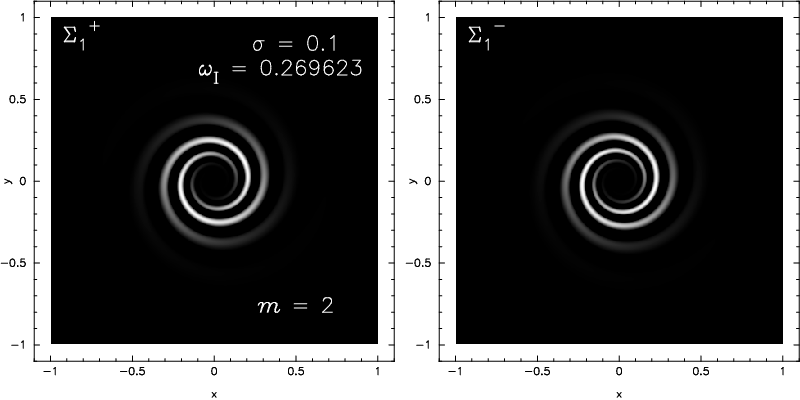}
\includegraphics[angle=0,scale=0.4]{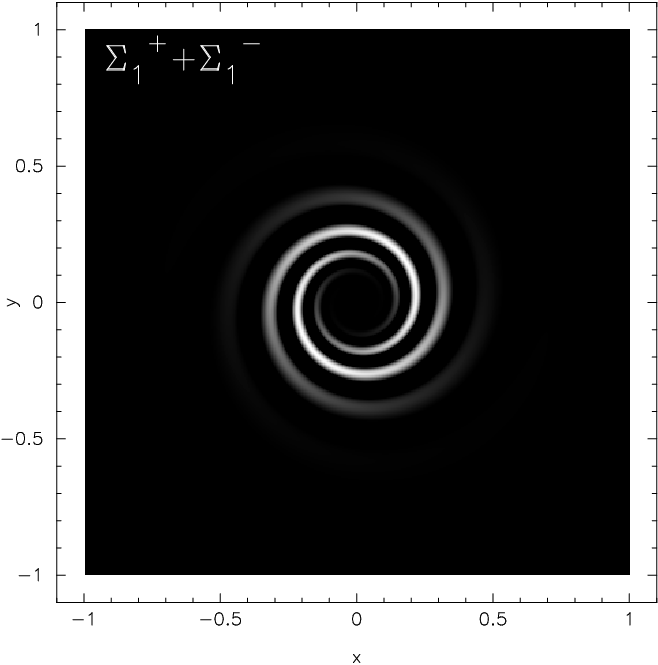}
\caption{This plot displays the gray-scale image of density enhancement regions 
due to the perturbations for Kuzmin disc profile in the $x-y$ plane for time $t = 0$. 
Plots are are labelled for their respective values of $\omega_{\text{\tiny{I}}}, 
m \,\,\&\,\, \sigma$. Wavepackets are more radially compact for lower values of 
$\sigma \,\,\& \,\, m$.}
\label{eigplot_3}
\end{center}
\end{figure}

Next, in fig.~\ref{eigplot_3} we plot the perturbed surface density in $x$-$y$ plane to 
show its variation as a function of $\sigma \,\,\& \,\,m$. First two rows display the positive 
component of real part of $\siga{1}{+}$, $\siga{1}{-}$ and $\siga{1}{+} + \siga{1}{-}$ 
for $m = 1, \sigma = 0.1 \,\,\&\,\, 0.2$ for the highest growth rate for each value of $\sigma$. 
As the velocity dispersion decreases---in other words for colder discs---the eigenmodes get 
radially more compact, although the radial extent is larger for colder discs. The last third 
row is the same plot for $\sigma = 0.1$ and $m = 2$. The modes are radially more compact for 
lower value of $m$. Apart from detailed structure of the eigenfunctions, the general properties 
of the eigenfunction remain the same for both chosen surface density profiles, and therefore 
we do not display the plots for JT annular disc to avoid repetition. Moreover, all the 
features we get here are consistent with the earlier works of \citet{st10, gss12}. However, 
the present approach has advantages as pointed out earlier.

\begin{table}
\centering
\begin{center}
\begin{tabular}{|c|c|c|c|} 
\hline
&&&\\
$\omega_{\text{\tiny{I}}} \downarrow$&$\sigma = 0.1$&$\sigma = 0.2$&$\sigma = 0.3$\\
&&&\\
\hline
&&&\\
1&0.4261042&0.2084615&0.1096090\\
&&&\\
2&0.4261249&0.2074471&0.1062024\\
&&&\\
3&0.3616852&0.1364123&0.0524338\\
&&&\\
4&0.3617675&0.1347891&0.0494986\\
&&&\\
5&0.3088861&0.0909412&0.0281082\\
&&&\\
6&0.30906861&0.0893201&0.0262204\\
\hline
\end{tabular}
\caption{This table gives the first six discrete eigenvalues (growth rate) 
for the case equal counter-rotation, for different 
values of sigma. These values are for Kuzmin disc profile and $m = 1$. Entries of row 
`$(1\,\,\&\,\,2)$', `$(3\,\,\&\,\,4)$' and `$(5\,\,\&\,\,6)$' forms degenerate pairs of 
eigenvalues.}
\label{table_1}
\end{center}
\end{table} 

\subsection{Other values of $\eta$}\label{othr_values}

In this subsection we present the results for values of $\eta$ other then 
$0 \,\,\& \,\,1/2$. For these values, the discrete spectrum of eigenvalues we get 
are complex with non-zero real and imaginary parts, and we write 
$\omega = \omega_{\text{\tiny{R}}} + \rmi\omega_{\text{\tiny{I}}}$. Such 
modes are interesting as these correspond to growing/damping modes with the 
growth rate given by $\omega_{\text{\tiny{I}}}$ which also precess with the pattern speed 
given by $\omega_{\text{\tiny{R}}}$. We use $\eta = 0.25 \,\,\& \,\,0.4$ as test cases. 

\begin{figure}
\begin{center}
\includegraphics[angle=0,scale=0.85]{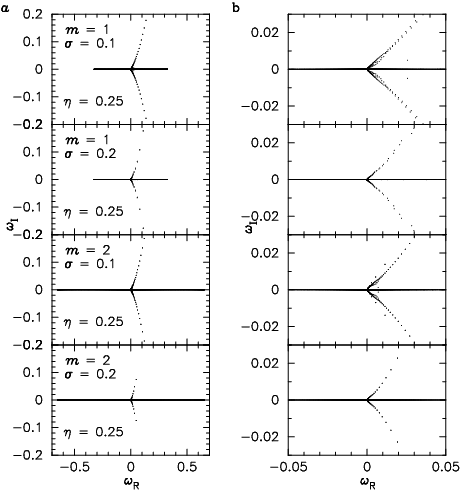}
\includegraphics[angle=0,scale=0.85]{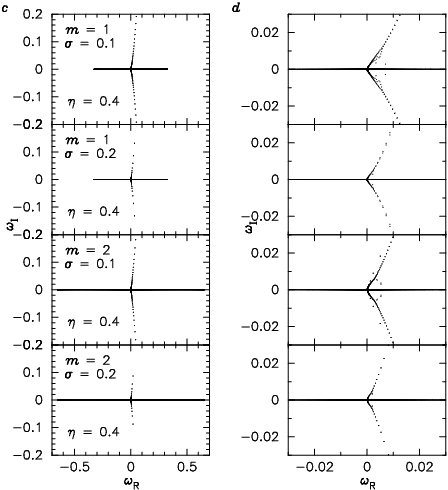}
\caption{Eigenvalue plot for $\eta = 0.25$ (panel $(a)$ and $(b)$) and $\eta = 0.4$ 
(panel $(c)$ and $(d)$) in complex Argand-plane for Kuzmin disc profile. panel $(a)$ 
and $(c)$ gives the whole spectrum whereas column $(b)$ and $(d)$ are the close-up view 
near origin of their corresponding plots on the left. panels are labelled for 
their respective values of $m \,\,\&\,\,\sigma$. Horizontal line gives the continuum 
of singular (van-Kampen) eigenmodes given by $\omega = -m\pom$. Discrete spectrum 
of eigenvalue forms a wedge like shape as the eigenvalues exists is complex conjugate 
pairs. Close view near the origin shows a double armed structure due to the presence 
of degenerate pairs of eigenvalues.}
\label{spectrum_1}
\end{center}
\end{figure}

\begin{figure}
\begin{center}
\includegraphics[angle=0,scale=0.85]{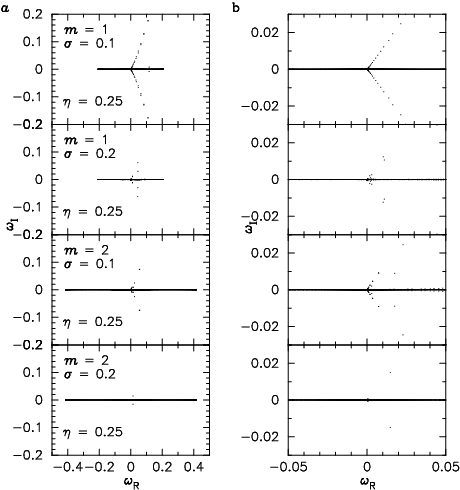}
\includegraphics[angle=0,scale=0.85]{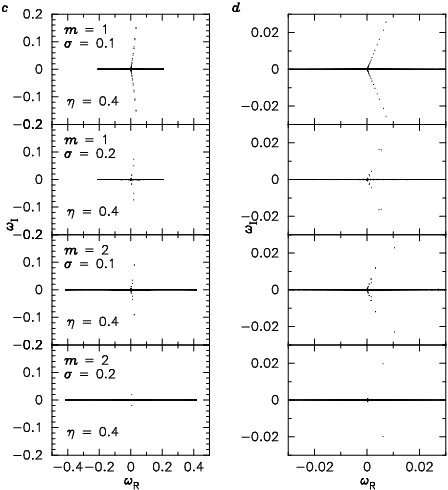}
\caption{Eigenvalue plot in complex argand-plane for JT annular disc on the same lines as that 
of fig.~\ref{spectrum_1}. Major difference in the spectrum we see is in the discrete 
part of the spectrum. The eigenvalues are sparsely spaced and we do not see the prominent double 
armed structure as seen for Kuzmin disc.}
\label{spectrum_2}
\end{center}
\end{figure}

In fig.~\ref{spectrum_1} and \ref{spectrum_2} we display the eigenvalues in 
complex argand plane. Fig.~\ref{spectrum_1} is for Kuzmin disc and 
Fig.~\ref{spectrum_2} is for JT annular disc. The panel labelled $(a)$ is 
for $\eta = 0.25$ and the panel labelled $(c)$ is for $\eta = 0.4$. First two rows 
are for $m = 1$ and $\sigma = 0.1 \,\,\& \,\,0.2$ and bottom two rows are for $m = 2$, 
for same values of velocity dispersion. The panel labelled $(b)$ and $(d)$ are the 
close-up view of regions near origin of the corresponding panels on the left. The 
horizontal lines are the real and continuous part of the spectrum. The continuum of 
eigenvalues corresponds to singular (van Kampen) modes. The eigenfunctions are concentrated 
at inner Lindblad resonances, which occurs at the radii for which $\omega = \pm m\pom$ 
(`$\pm$' signs are for prograde and retrograde discs respectively). Note that since for 
slow modes $\omega \ll \Omega$, the corotation and outer Lindblad resonances do not exist 
as is also pointed out by \citet{gss12}. Both the surface density profiles 
display the same behaviour for the continuous part of the eigenspectrum.

Coming to complex eigenvalues (or the discrete part of the spectrum), 
we get a wedge-like distribution, as the eigenvalues exist in complex conjugate pairs. 
For $\eta = 0$ the eigenvalues are purely real, and 
as we increase the value of $\eta$ the spectrum goes from real to complex, until 
for $\eta = 1/2$ the discrete spectrum is purely imaginary. This transition was first 
found by \citet{tou02}, and later by \citet{gss12}\footnote{\citet{tou02} attributes 
this bifurcation to a phenomenon identified by M. J. Krein due to resonant crossing 
of stable modes.}. A close-up view in panel $(b)\,\,\&\,\,(d)$ in fig.~\ref{spectrum_1} 
we see that these two branches consists of more than one arm. These arms are due to 
the presence of degenerate pairs of eigenvalues corresponding to leading and trailing waves. 
Such pairs exists all throughout the branch. The separation in the 
degenerate pairs increases as we go to lower values of eigenvalues. 
Hence the arms separate out more prominently close to origin and go to 
zero with further decrease of eigenvalue. These arms were also noticed by \citet{gss12} 
while studying the softened gravity disc. We do not see such prominent double-armed 
structure in the case of JT annular disc. This could probably be due to the fact that 
discrete spectrum itself is quite sparse. Secondly, in the case of JT annular disc, 
most of the disc mass is concentrated in an annular region, and as pointed by \citet{jt12} 
the degenerate pair of eigenvalues merge if the disc mass at resonances shrinks to zero. 
\citet{tou02} have also studied softened gravity counter-rotating disc, applicable to 
planetary discs. The spectrum of eigenvalues the authors get is scattered and the 
plausible reason for this could be that the presence of degenerate pairs and the arms are 
not well defined due to sparse nature of the eigenvalues. For the variation with 
$m \,\,\& \,\,\sigma$, the largest growth rate is a decreasing function of $\sigma$ 
and $m$ both.

\begin{figure}
\begin{center}
\includegraphics[angle=0,scale=0.665]{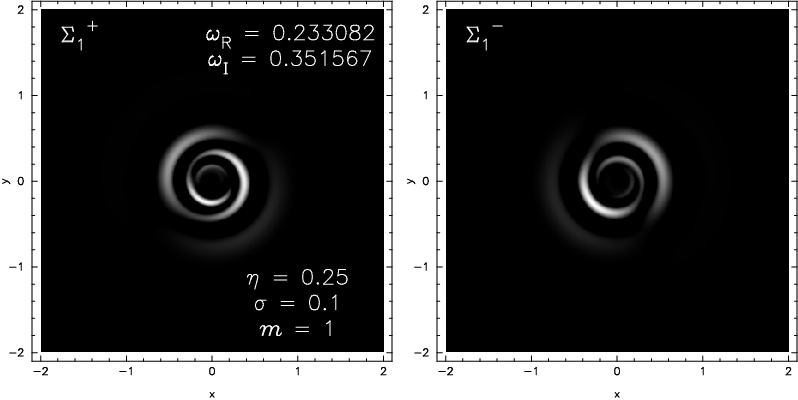}
\includegraphics[angle=0,scale=0.4]{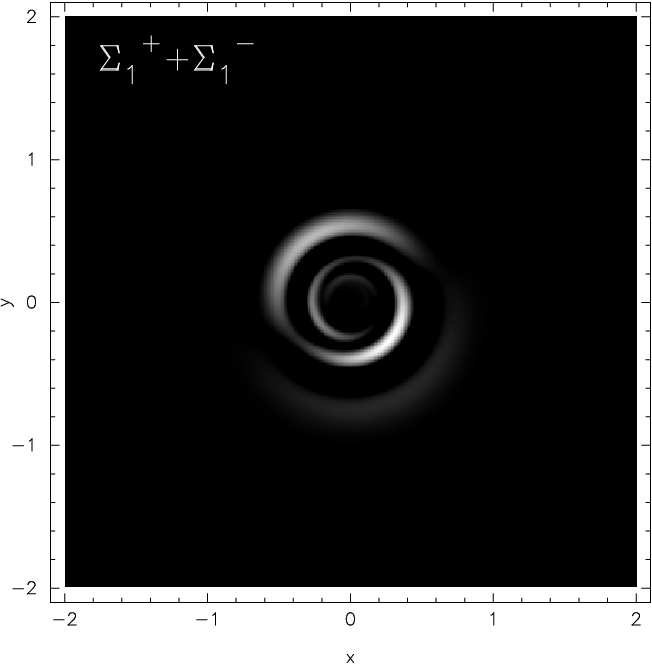}
\includegraphics[angle=0,scale=0.665]{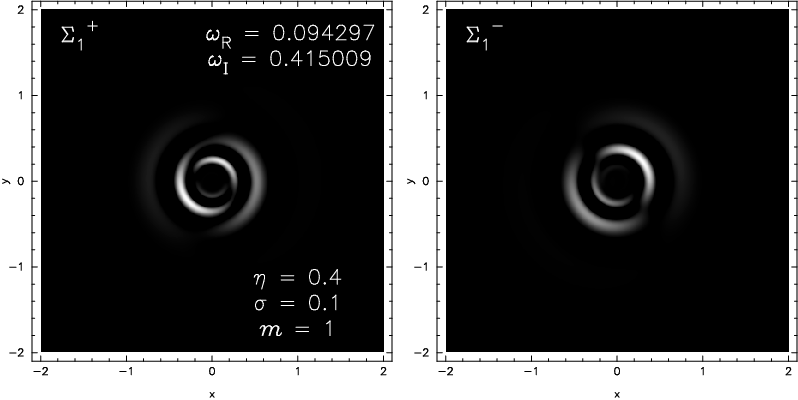}
\includegraphics[angle=0,scale=0.4]{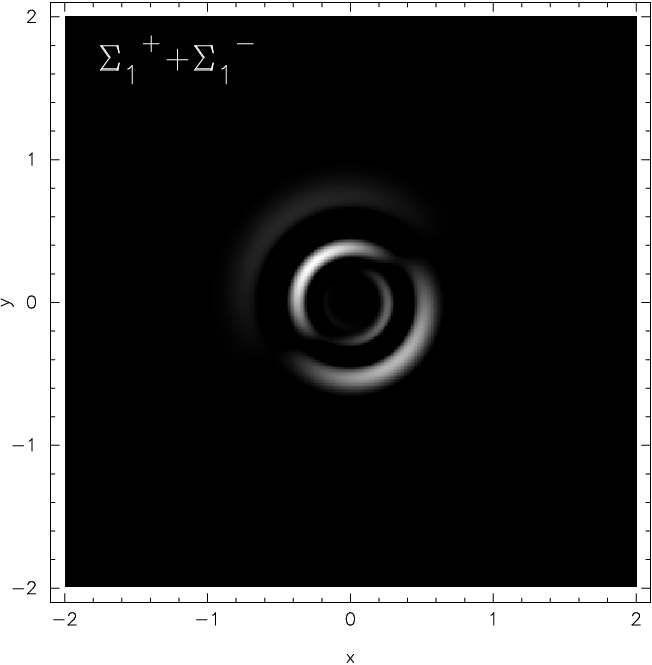}
\caption{Gray-scale plot of the positive component of real part of 
$\siga{a}{\pm}(R)\exp(\rmi(m\phi - \omega t))$ and $\siga{a}{+}(R)\exp(\rmi(m\phi - \omega t)) + 
\siga{a}{-}(R)\exp(\rmi(m\phi - \omega t))$ at time $t = 0$ for $\eta = 0.25 \,\,\& \,\, 0.4$ in 
top and bottom panel respectively. First column is labelled for their respective values of 
$m,\,\,\sigma\,\,\&\,\,\omega$. Lopsidedness in the density profile is evident and is more 
prominent for higher value of $\eta$.}
\label{eigplot_4}
\end{center}
\end{figure}

\begin{figure}
\begin{center}
\includegraphics[angle=0,scale=0.9]{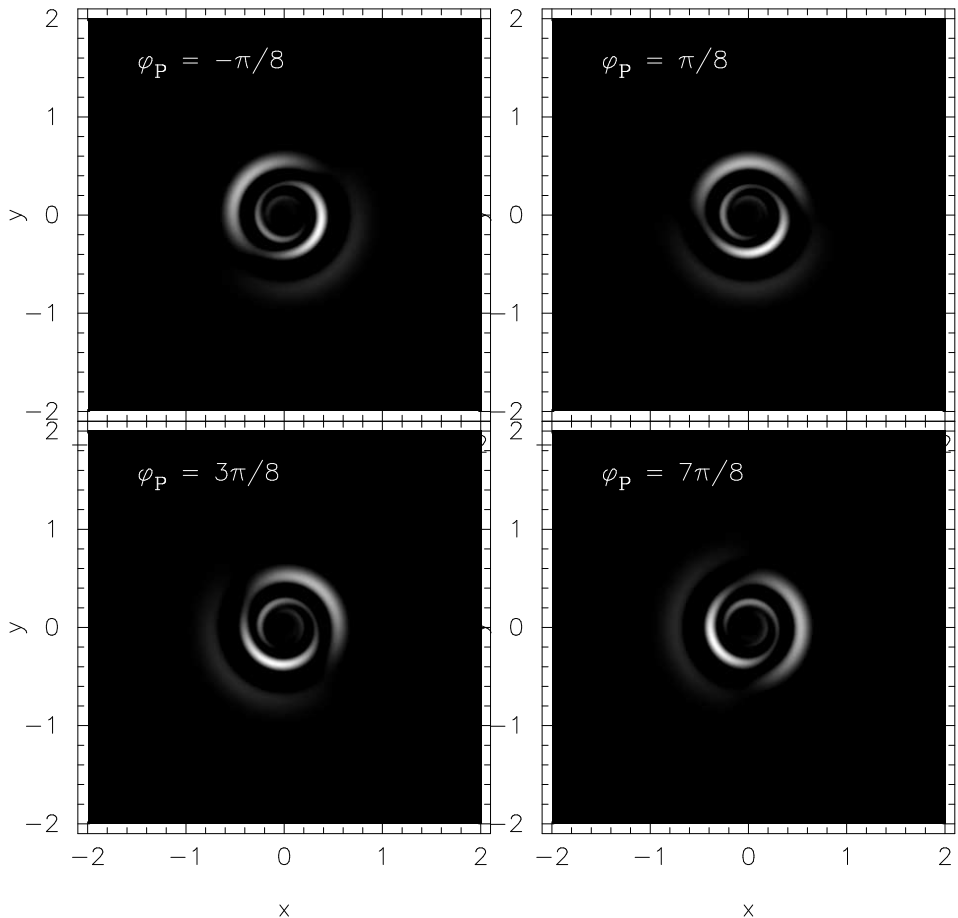}
\caption{This plot displays the time evolution of total perturbed surface density 
for the same model parameters same as that in top panel of fig.~\ref{eigplot_4}. The 
plots are labelled for the value of $\varphi_P = \omega_{\text{\tiny{R}}} t$. Pattern 
rotates with angular speed given by $\omega_{\text{\tiny{R}}}$ along with an exponential 
increase in the intensity.}
\label{eigplot_5}
\end{center}
\end{figure}

Eigenfunctions are in general complex. Fig.~\ref{eigplot_4} is the 
gray-scale image of the positive component of the real part of the perturbed 
surface density, $\siga{1}{\pm}$ and $\siga{1}{+} + \siga{1}{-}$, for the Kuzmin 
disc as the unperturbed disc profile, $m = 1$ and $\sigma = 0.1$. Top row is for 
$\eta = 0.25$ and $\omega = 0.233082 + \rmi 0.351567$ whereas the second row is 
for $\eta = 0.4$ and $\omega = 0.094297 + \rmi 0.415009$. The density contrast is 
clearly lopsided, which is more prominent for higher values of $\eta$. In 
fig.~\ref{eigplot_5} we display the snapshots of evolution of positive part of total 
surface density perturbation, where the panels are labelled for 
$\varphi_P = \omega_{\text{\tiny{R}}} t$ for the same parameters as that of 
top panel of fig.~\ref{eigplot_4}. Each one contains certain amount of lopsidedness. 
The pattern rotates at an angular speed given by $\omega_{\text{\tiny{R}}}$ and 
there is an overall increase in the magnitude as the system evolves (exponential 
increase in magnitude because of the presence of $\exp(\omega_{\text{\tiny{I}}} t)$).

\section{Summary and conclusions}\label{cnclsns}
We have formulated and analysed the modal behaviour of a system of two nearly Keplerian, 
counter-rotating axisymmetric stellar discs, rotating around a central mass. The formalism 
is a generalisation of the one studied in Paper~I for a single disc, where we go one step 
beyond the usual WKB analysis by not assuming the relation between perturbed potential 
and surface density to be local. We first derived the integral eigenvalue equation for a 
tightly-wound linear modes of coplanar axisymmetric counter-rotating discs under the epicyclic 
approximation. Then, as an application of this equation, we restricted ourselves to near 
Keplerian systems---which support slow modes. We took the local limit of the integral equation 
to obtain the WKB dispersion relation to study the stability of the discs and concluded that 
(i) Counter-rotating discs are {\it stable} to axisymmetric perturbations if they satisfy 
the well known Toomre stability criterion. (ii) Non-axisymmetric perturbations are stable for 
a single disc, consistent with the conclusions of \citet{st10,gss12,jt12} and Paper~I. 
(iii) For non-zero mass in retrograde disc, discs are found to be largely unstable to 
non-axisymmetric perturbations.

Next we solved the integral eigenvalue equation for slow modes numerically. We used two 
different unperturbed surface density profiles, namely, Kuzmin disc, which is a centrally 
concentrated disc profile; and JT annular disc, which is a annular disc introduced by 
\citet{jt12}. We assumed for both `$\pm$' discs the same radial profile of velocity dispersion 
$\sigma_R = \sigma R\kappa(R)$ with $\sigma < 1$. The same profile is  also used in Paper~I for 
a single disc, which was motivated by the work of \citet{jt12}. We solved for various values 
of mass fraction, $\eta$ in the retrograde disc for $m = 1 \,\&\, 2$. The spectrum for $\eta = 0$ 
served as a test for our numerical methods, and the results obtained are in exact correspondence 
with those in Paper~I. Following are the general properties of the spectrum and eigenfunctions 
obtained by us for all values of $\eta$.
\begin{itemize}
 \item  At $\eta = 0$ the eigenvalues are all real and the modes are 
 stable and oscillatory. As we increase the value of $\eta$, the eigenvalues become 
 imaginary, with the value of highest growth rate increasing with increasing mass 
 fraction in the retrograde disc till $\eta = 1/2$. As we further increase $\eta$ the 
 value of largest growth rate declines and the spectrum is again  purely real (stable 
 modes) for $\eta = 1$.
 \item For no counter--rotation all the trends (discussed in the beginning of 
 $\S~\ref{num_rslts}$) in the eigenspectrum and waveforms favour the observational 
 detection of eigenmodes with lower values $m\,\,\&\,\,\sigma$. 
 \item For equal counter--rotation, the highest growth rate for a given set of parameters 
 decreases as a function of $\sigma\,\,\&\,\,m$, also favouring the excitation and hence 
 detection of modes with lower values  $\sigma\,\,\&\,\,m$.
 \item For other values of $\eta$, both real and imaginary parts of the eigenvalues 
 are non-zero. The real part gives the pattern speed of the eigenfunction and 
 imaginary part gives its growth rate.
 \item Eigenvalues exist in degenerate pairs, corresponding to leading and trailing 
 waves, for all values of $\eta$. The presence of such degenerate pairs explains the 
 presence of double-armed structure as seen the spectrum of eigenvalues for non-zero 
 counter--rotation. Such double armed structure were also seen in \citet{gss12}. 
 \item The separation between the degenerate pairs of eigenvalues increases with 
 increasing $\sigma$ values and increasing number of nodes in the eigenfunction.
 \item The plot for surface density enhancement in the disc plane shows an overall 
 lopsided intensity distribution for $m= 1$, which is more prominent for higher values of $\eta$. 
 \item The growth rate of pattern increases for higher values of $\eta$ and lower values of $m$, 
 therefore allowing the $m = 1$ instabilities to play a dominant role in the dynamics of such systems. 
\end{itemize}

The calculated waveforms and their properties favour that these eccentric modes are the 
behind various non-axisymmetric features seen the discs like galactic nuclei, debris 
discs and accretion discs around stellar mass compact objects. Presence of retrograde 
mass in the disc gives rise to instabilities, thereby generating long lived, large 
scale features. A natural question to ask at this stage is how these counter--rotating 
streams of matter are formed in discs. In case of galaxies like M$31$, such retrograde 
orbits could be the result of in-fall of debris into the centres of such galaxies. 
\citet{ss02} proposed that these stars could have been accreted to the centre of M31 
in the form of a globular cluster that spiralled-in due to dynamical friction. For 
the in-falling mass in galactic nuclei, the sense of rotation will be uncorrelated with 
respect to the pre-existing material. Thus in the course of evolution of a galaxy, it is 
probable that counter-rotating systems are generically produced. Simulations by \citet{nix12} 
show breaking of disc in the vicinity of rapidly rotating central massive object, and hence 
generically forming counter--rotating discs. Such counter--rotating streams of matter are 
also thought to be helpful in feeding massive black holes at centres of many galaxies.

Present work as well as the previous studies on slow modes for galactic discs are done 
assuming the disc to be composed of zero pressure fluid (gas) disc \citep{tre01,ss02,gss12} 
or collisionless disc composed of stars (\citet{jt12}, Paper~I). However in case of galaxies 
the gas/dust and the stellar discs exists together and are coupled to each other. In future 
we aim to formulate a theory of linear eigenmodes for discs composed of both stars and dust 
to study the nature of modes if both gas and stellar discs interact with each other only 
gravitationally. Recently \citet{jalali13} have done a similar problem applicable only to 
protoplanetary discs using numerical simulations and found such systems to be unstable. 
We plan to extend our semi-analytical formulation to study the linear perturbation theory 
for such discs.

\end{document}